\documentclass[aps,prl,reprint,10pt,twocolumn]{revtex4-1}
\usepackage{graphicx}
\usepackage{epsfig}
\usepackage{amssymb}
\usepackage{amsmath}
\usepackage{amsfonts}
\usepackage{color}
\usepackage[latin1]{inputenc}
\usepackage{float}
\usepackage{comment}

\begin{document}
\title{Universality and dependence on initial conditions in the class of the nonlinear molecular beam epitaxy equation}
\author{I. S. S. Carrasco}
\author{T. J. Oliveira}
\email{tiago@ufv.br}
\affiliation{Departamento de F\'isica, Universidade Federal de Vi\c cosa, 36570-900, Vi\c cosa, Minas Gerais, Brazil}

\begin{abstract}
We report extensive numerical simulations of growth models belonging to the nonlinear molecular beam epitaxy (nMBE) class, on \textit{flat} (fixed-size) and \textit{expanding} substrates (ES). In both $d=1+1$ and $2+1$, we find that growth regime height distributions (HDs), and spatial and temporal covariances are universal, but are dependent on the initial conditions, while the critical exponents are the same for flat and ES systems. Thus, the nMBE class does split into subclasses, as does the Kardar-Parisi-Zhang (KPZ) class. Applying the ``KPZ ansatz'' to nMBE models, we estimate the cumulants of the $1+1$ HDs. Spatial covariance for the flat subclass is hallmarked by a minimum, which is not present in the ES one. Temporal correlations are shown to decay following well-known conjectures.
\end{abstract}


\maketitle


Scaling invariance and universality, two pillars of the theory of phase transitions and critical phenomena, have been also very important in the study of nonequilibrium systems \cite{OdorBook}. One of the most prominent examples is the dynamics of growing interfaces, whose width $w(L,t)$ increases in time as $w \sim t^{\beta}$ (while the correlation length $\xi$ parallel to the substrate scales as $\xi \sim t^{1/z}$), and with the system size as $w \sim L^{\alpha}$ (when $\xi \sim L$). A set of exponents $\alpha$, $\beta$ and $z$ defines an universality class and, interestingly, only a few classes exist, which are determined by some fundamental symmetries \cite{Barabasi}. For example, interfaces evolving under tension and growing in the direction of its local normal are expected to belong to the Kardar-Parisi-Zhang (KPZ) class, being described at a coarse-grained level by the KPZ equation \cite{kpz}
\begin{equation}
\frac{\partial h}{\partial t} = F + \nu_2 \nabla^2 h + \frac{\lambda_2}{2} (\nabla h)^2 + \zeta(\vec{x},t).
\label{eqKPZ}
\end{equation}
The Edwards-Wilkinson (EW) \cite{EW} equation (class) is given by $\lambda_2=0$. On the other hand, when the growth is dominated by the surface diffusion of adatoms, as is the case in molecular beam epitaxy (MBE), it is expected to fall into the nonlinear MBE (nMBE) class, associated with the equation by Villain \cite{Villain} and Lai and Das Sarma \cite{LDS}
\begin{equation}
\frac{\partial h}{\partial t} = F - \nu_4 \nabla^4 h + \lambda_4 \nabla^2 (\nabla h)^2 + \zeta(\vec{x},t),
\label{eqVLDS}
\end{equation}
or in its linear counterpart (with $\lambda_4=0$). In all these growth equations, $h(\vec{x},t)$ is the height at substrate position $\vec{x}$ and time $t$; $F$, $\nu_i$ and $\lambda_i$, with $i=2,4$, are constants and $\zeta(x,t)$ is a white-noise, with $\left\langle \zeta \right\rangle = 0$ and variance $\left\langle \zeta(x,t)\zeta(x',t') \right\rangle = 2 D \delta^{d_s}(x-x')\delta(t-t')$ \cite{Barabasi}.

Recent theoretical \cite{SasaSpo1,*Amir,*Calabrese,*Imamura}, experimental \cite{TakeuchiPRL,*TakeuchiSP} and numerical \cite{[{For recent survey of literature see, e.g.,}]healytake2015} works on KPZ systems have changed our view of KPZ universality by demonstrating that this class splits into subclasses depending on initial conditions (ICs), or surface geometry. More specifically, while the scaling exponents ($\alpha$, $\beta$ and $z$) are the same for KPZ growth starting from a flat substrate (\textit{flat} IC/geometry) or from a seed - so that the active growing zone expands in time - (usually called \textit{curved} or droplet geometry), the (1-point) height distributions (HDs) and (2-point) spatial and temporal correlators are different, but universal in each IC/geometry. In $d=1+1$, the height fluctuations are given by Tracy-Widom distributions \cite{TW}, and spatial covariances are associated with Airy processes \cite{PraSpo3,*Sasa2005,*Borodin.etal-CPAM2008}. In higher dimensions, universality and IC dependence of KPZ HDs have been demonstrated numerically \cite{healy12,*Oliveira13,Alves14,healytake2015} and confirmed experimentally for the ($2+1$) flat subclass \cite{Almeida14,*healy2014,*Almeida15}.

Despite the importance of the nMBE class - since MBE is the main technique for thin film deposition - basically nothing is known about universality of (growth regime) HDs and IC dependence in these systems. In order to decrease this abyss between KPZ and nMBE classes, in this work we present a detailed numerical analysis of nMBE models studied on flat substrates of fixed-size (flat IC) and enlarging sizes (ES IC, which mimic the curved geometry \cite{Ismael14}). Results from large scale simulations, in $d=1+1$ and $2+1$, demonstrate that universal and IC-dependent HDs and correlators also exist in nMBE growth. Beyond the obvious application of the flat subclass (for MBE growth on flat substrates), we note that the ES one might be appealing for deposition on textured substrates. A prominent example, which is very important for several applications \cite{Savin15}, is etched Si(100) surfaces, where inverted pyramid holes can be formed \cite{Wang15} and, depending on the growth conditions and Si-adsorbate affinity, expanding surfaces might be observed while growth proceeds inside the holes.

To distill the universality of HDs, let us consider the so-called ``KPZ ansatz" \cite{PraSpo1}
\begin{equation}
 h = v_{\infty} t + (\Gamma t)^{\beta} \chi + \eta + \ldots,
\label{eqansatz}
\end{equation}
where $v_{\infty}$ (the asymptotic growth velocity), $\Gamma$ (setting the amplitude of $w$) and $\eta$ (a stochastic correction) are non-universal (system-dependent) parameters, while $\chi$ is a random variable yielding the height fluctuations (which are universal in the KPZ class). A simple analysis of Eq. \ref{eqVLDS}, considering periodic boundary conditions (PBC), shows that the mean height is always given by $\left\langle h \right\rangle = F t$. Comparing this with Eq. \ref{eqansatz}, one sees that $v_{\infty}$ is equal to the deposition flux ($v_{\infty}=F$), while the mean of the nMBE HDs is null (i. e., $\left\langle \chi \right\rangle = 0$), as well as are corrections in $\left\langle h \right\rangle$. This implies that the shift observed in the mean of KPZ HDs does not exist in the nMBE ones, since $\left\langle \eta \right\rangle = 0$. The exponents $\alpha = (4 - d_s)/3-\delta$ and  $z = (8 + d_s)/3-2\delta$, and so $\beta=\alpha/z$, are exactly known from two-loop renormalization, where $\delta=0.01361(2-d_s/2)^2$ is a correction to the one-loop result \cite{Janssen}. Following a dimensional analysis of the nMBE equation, as done in Ref. \cite{AmarFamily1,*AmarFamily2} for 1-loop exponents ($\delta=0$), we find here the scaling of the variance of HDs (for 2-loops) as
\begin{equation}
 \left\langle h^2 \right\rangle_c = w_2(L,t) = A L^{2 \alpha} f[(\xi(t) / L)^{z}],
 \label{eqAF}
\end{equation}
where $\xi(t) = (D A^{-1} t)^{1/z}$ is the correlation length and $A = (D/\lambda_4)^{2/3} [\nu_4^3/(\lambda_4^2 D)]^{2\delta/(4-d_s)}$ sets the roughness amplitude at the steady state regime (where $f(x) \sim const$). In the growth regime, $f(x) \simeq b x^{2 \beta}$, so that $w_2(\infty,t) = b [D A^{\frac{1}{2\beta} -1 } t]^{2 \beta}$. Comparing this with Eq. \ref{eqansatz}, one may identify $\Gamma = D A^{\frac{1}{2\beta} -1 }$ and $b = \left\langle \chi^2 \right\rangle_c$.


The standard discrete model in the nMBE class is the conserved restricted solid-on-solid (CRSOS) model \cite{CRSOS1}, where a (randomly deposited) particle aggregates in a site $i$ (i. e., $h_i \rightarrow h_i+1$) if the restriction $|h_{i}-h_{j}| \leq m$ is satisfied for all nearest-neighbors (NN) $j$. Otherwise, it is deposited at the nearest site of $i$ satisfying the restriction \cite{CRSOS1}. Theoretical calculations \cite{ParkKimPark1,*ParkKimPark2} for this model with $m=1$ (hereafter called CRSOS1), in $d=1+1$, have demonstrated that it is described by the nMBE equation, in the hydrodynamic limit, with parameters $\nu_4=(21-12\sqrt{2})/2$, $\lambda_4=(10-3\sqrt{2})/2$, and $D=(2\sqrt{2}-1)/2$. Therefore, $A=0.4662$ and $\Gamma=0.6167$ for this model~\footnote{The one-loop ($\delta=0$) values are $A=0.4655$ and $\Gamma=0.6237$, differing only slightly from the two-loop ones.}, which will be used as a benchmark in our analyses. Another classical nMBE model is the one from Das Sarma and Tamborenea (DT) \cite{DT}, where the freshly (randomly) deposited particle, in a site $i$, can move to its NN sites in order to increase the number of lateral neighbors. While the scaling of the original DT model is featured by strong corrections, a version with noise reduction, where an aggregation occurs at a given site $i$ only after $N$ deposition is attempted at that site, displays scaling exponents in good agreement with the nMBE class in $d=1+1$ \cite{DTnr}. Data for $N=20$ are presented in the following \footnote{We have verified that simulations for $N \in [10,100]$ yield the same asymptotic results.}. Extensive simulations of the CRSOS model on substrates of \textit{fixed} lateral sizes up to $L=2^{17}$ ($d=1+1$) and $L=2^{12}$ ($2+1$) were carried out for $m=1$, $2$ (CRSOS2) and $4$ (CRSOS4). The DT model is investigated in $d=1+1$ for the same sizes. Furthermore, these models are also studied on enlarging substrates, using the method introduced by us in Ref. \cite{Ismael14}. In this case, the growth starts on (flat) substrates of lateral size $L_0=v_d$, which expand (in each dimension) at a constant rate $v_d$ by randomly duplicating columns. Here, one sets $v_d=12$ in $d=1+1$ and $v_d=1/2$ and $2$ in $2+1$. In all models, PBCs are considered, and the deposition flux is defined as one particle per site per time unit, so that $v_{\infty}=F=1$.


\begin{figure}[!t]
\centering
\includegraphics[height=3.5cm]{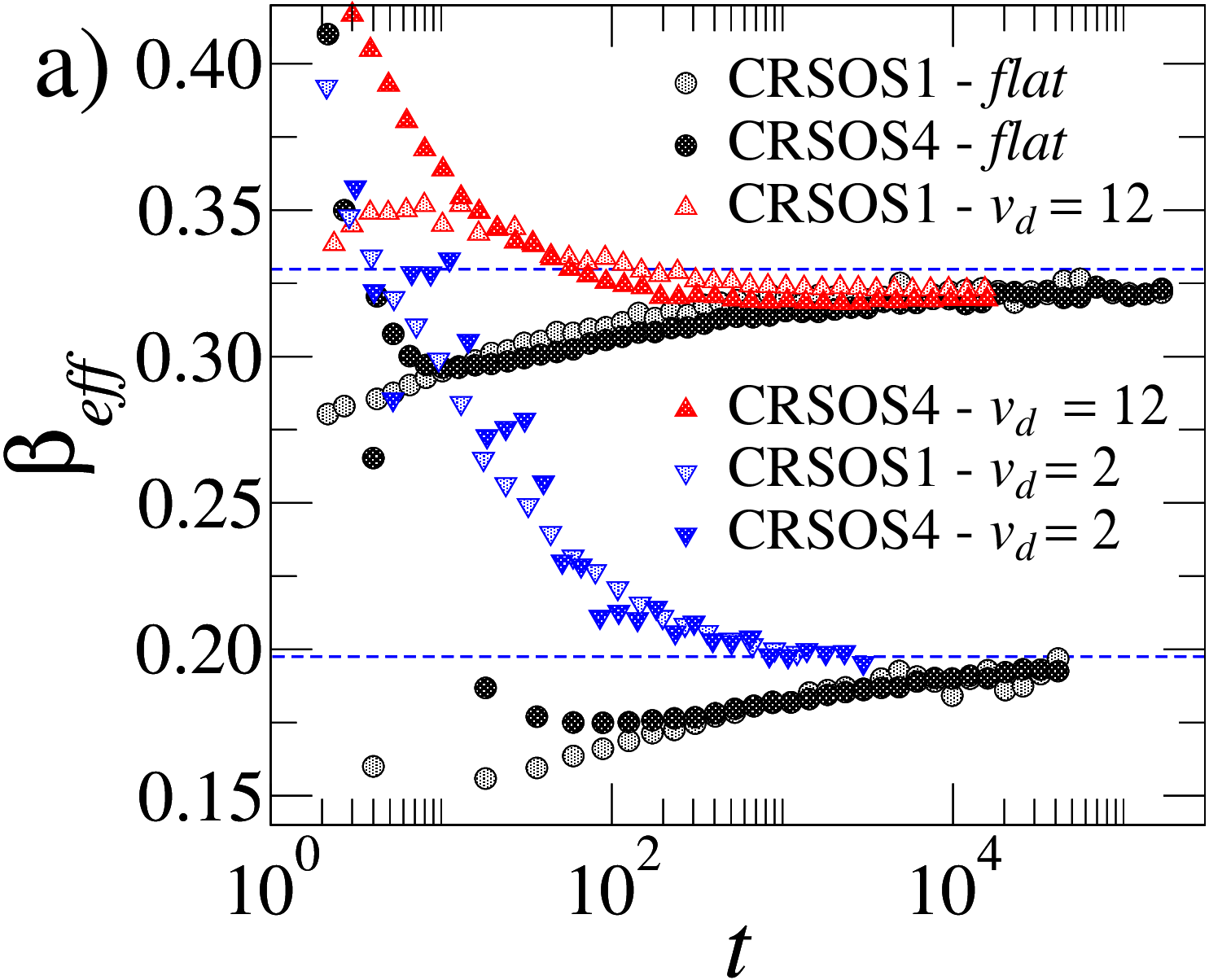}
\includegraphics[height=3.5cm]{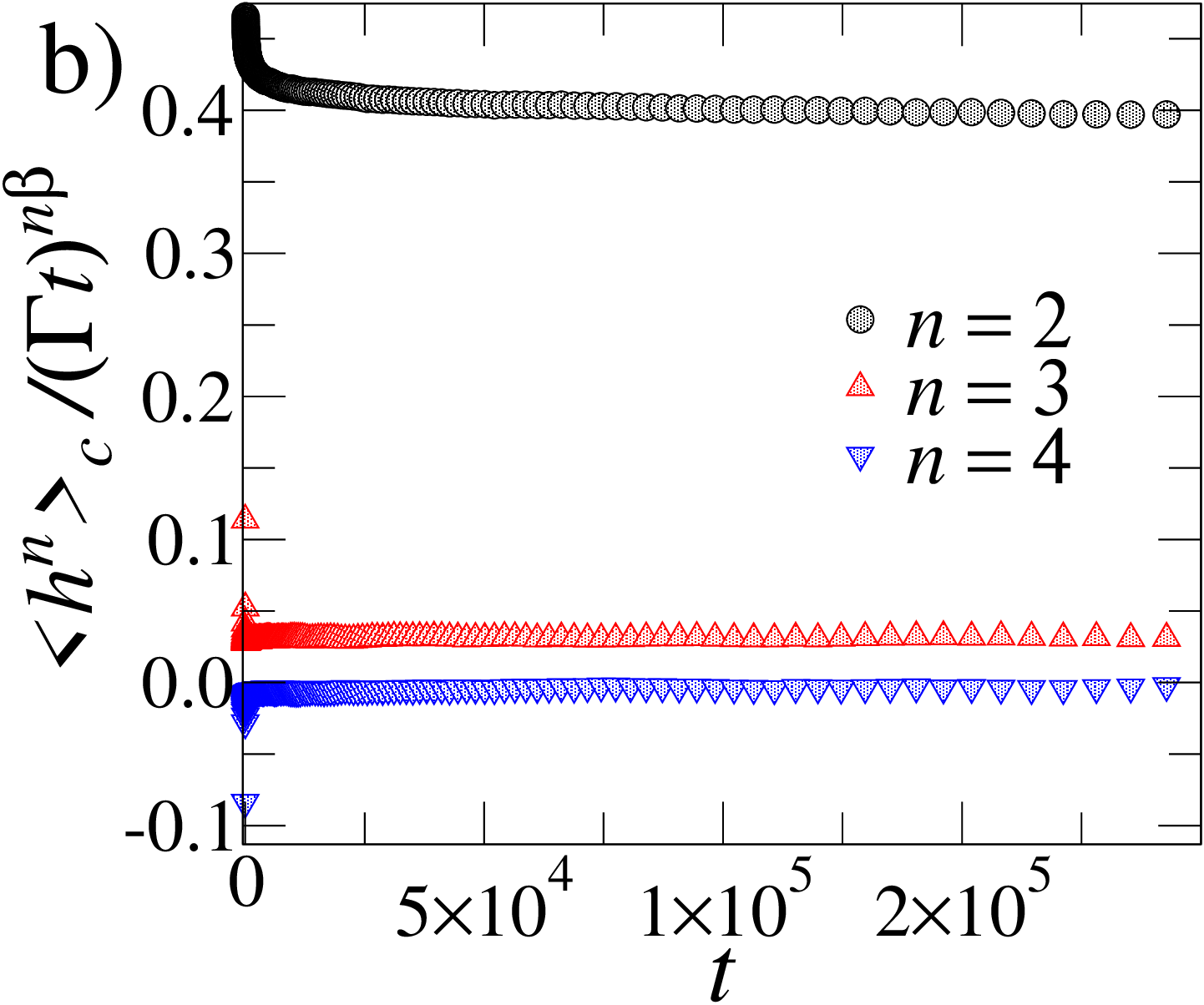}
\includegraphics[height=3.8cm]{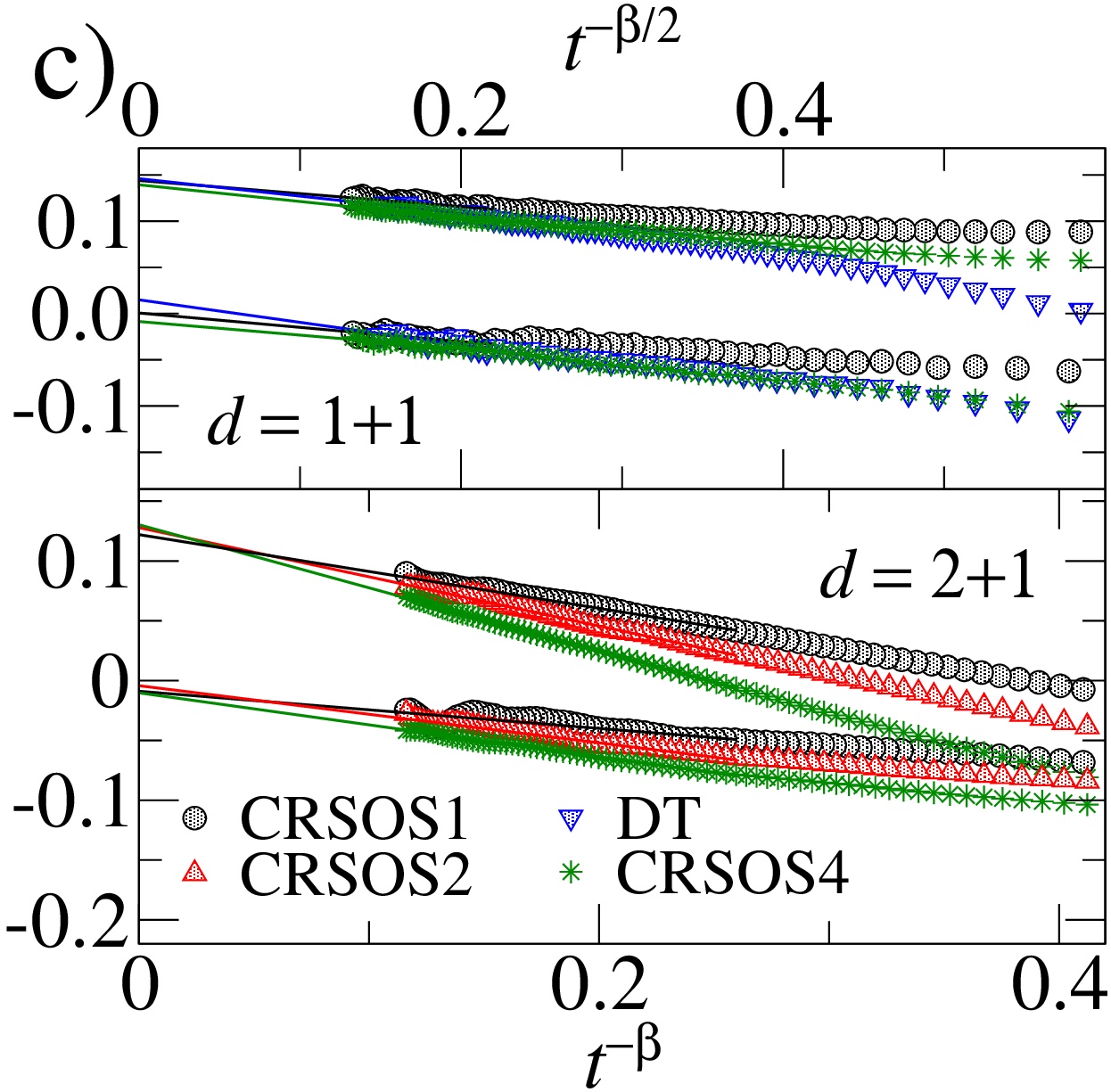}
\includegraphics[height=3.35cm]{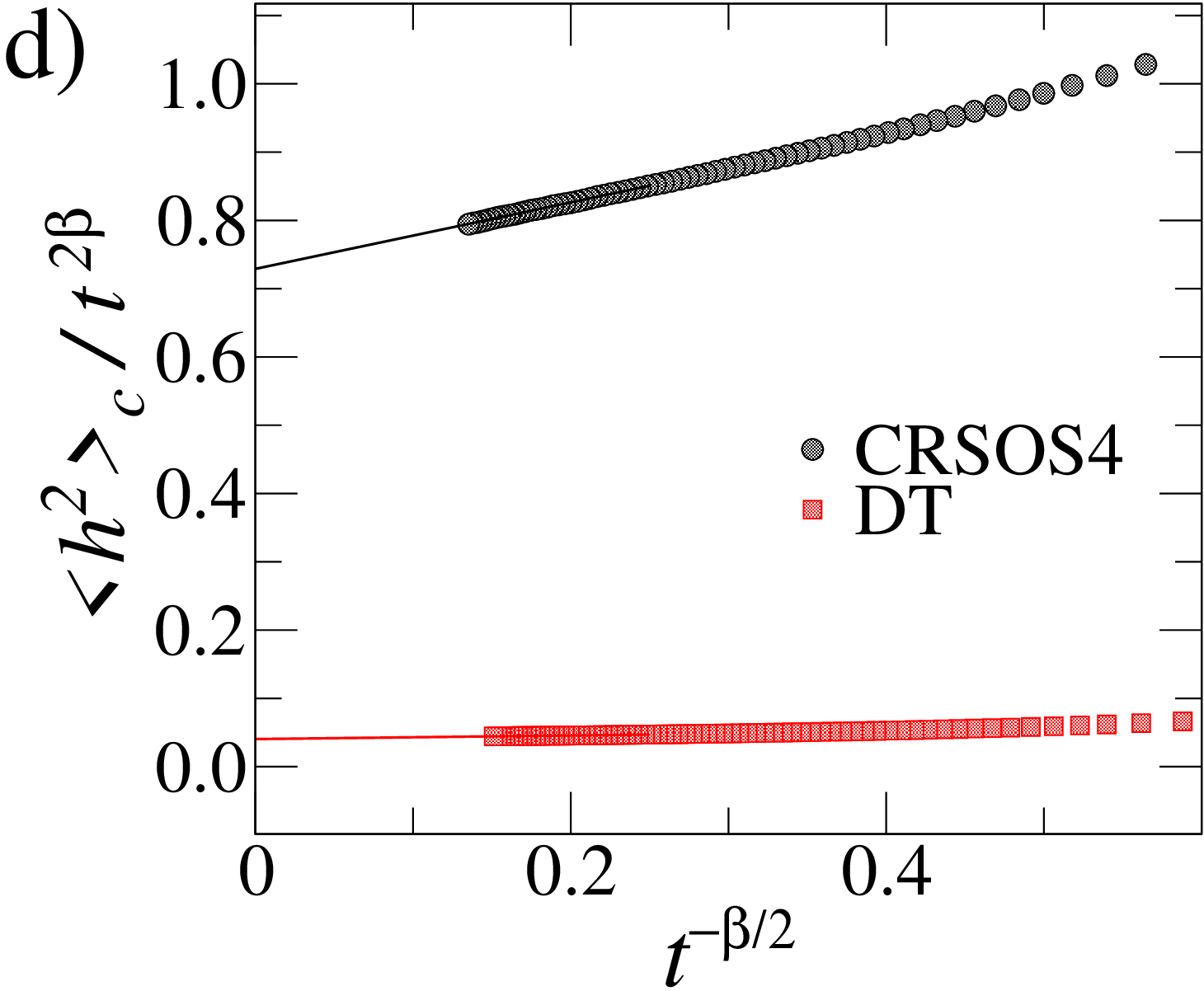}
\caption{(Color online) a) Convergence of effective growth exponents $\beta_{eff} \equiv \frac{1}{2}\frac{d \left( \ln \left\langle h^2 \right\rangle_c \right) }{d \left( \ln t \right) }$. b) Estimates of the cumulants of $\chi$ from $\left\langle h^n \right\rangle_c/(\Gamma t)^{n\beta}$ for the $1+1$ CRSOS1 model. c) Extrapolation of skewness (higher) and kurtosis (lower values) for $d=1+1$ (top) and $2+1$ (bottom panel). d) Estimates of $g_2$ from $\left\langle h^2 \right\rangle_c/t^{2\beta}$. All data in b)-e) are for flat IC.}
\label{fig1}
\end{figure}

\begin{table}[!b]
\caption{Asymptotic estimates of the first four cumulants of the HDs for the CRSOS1 model in $d=1+1$.}
\begin{center}
\begin{tabular}{c c c c c c c c c}
\hline \hline
IC & & $\langle \chi\rangle$ & & $\langle \chi^2\rangle_c$ & & $\langle \chi^3\rangle_c$ & & $\langle \chi^4\rangle_c$ \\
\hline
Flat    & &   0   & &   0.375(5)   & &  0.0315(5)   & &  0.000(2)    \\
ES    & &   0   & &   0.612(8)   & &  0.0487(3)   & &   0.001(3)   \\
\hline\hline
\end{tabular}
\label{tab0}
\end{center}
\end{table}

Effective growth exponents for CRSOS models with flat and ES ICs are compared in Fig. \ref{fig1}a. The convergence to the same asymptotic value demonstrates that the substrate enlargement preserves the roughness scaling properties, as expected \cite{Ismael14,Masoudi1,*Escudero2}. In general, one observes a very slow convergence of $\beta_{eff}$, which is still a bit smaller than the two-loop exponent even after very long times. This suggests the existence of additional corrections in the ansatz, so that $ h  = t + (\Gamma t)^{\beta} \chi + \mu t^{\varepsilon} + \dots$, where $\left\langle \mu \right\rangle = 0$, but $\left\langle \mu^2 \right\rangle_c \neq 0$ and/or the covariance $\left\langle \chi \mu \right\rangle_{cov} \neq 0$. Hence, $\left\langle h^2 \right\rangle_c/(\Gamma t)^{2\beta} = \left\langle \chi^2 \right\rangle_c + \Gamma^{-\beta} \left\langle \chi \mu \right\rangle_{cov} t^{\varepsilon-\beta} + \Gamma^{-2\beta} \left\langle \mu^2 \right\rangle_c t^{2(\varepsilon-\beta)} + \ldots$. Indeed, by plotting $\left\langle h^2 \right\rangle_c / (\Gamma t)^{2\beta}$ versus time for the $1+1$ CRSOS1 model (see Fig. \ref{fig1}b), instead of a constant $\left( \left\langle \chi^2 \right\rangle_c \right)$ one finds a slightly decreasing behavior consistent with $\left\langle \chi^2 \right\rangle_c + c t^{-\beta/2}$, so that $\varepsilon=\beta/2$ if $\left\langle \chi \mu \right\rangle_{cov} \neq 0$ or $\epsilon=3\beta/4$ otherwise. In any case, the extrapolation of $\left\langle h^2 \right\rangle_c / (\Gamma t)^{2\beta}$ to $t \rightarrow \infty$ give us the variance of the HDs (for the CRSOS1 model). Higher order cumulants are determined in the same way, from $\left\langle h^n \right\rangle_c/(\Gamma t)^{n\beta} = \left\langle \chi^n \right\rangle_{c} + \ldots$, as shown in Fig. \ref{fig1}b, for $n=3$ and $4$. The asymptotic cumulants for both ICs are summarized in Tab. \ref{tab0}. While $\langle \chi^4 \rangle_c \approx 0$ in both cases, mild and considerable differences exist in $\langle \chi^3 \rangle_c$ and $\langle \chi^2 \rangle_c$, respectively, demonstrating that the HDs are IC-dependent. In our analysis, we are assuming that $\Gamma$ is the same for flat and ES ICs \cite{Ismael14}.

\begin{table}[!t]
\caption{Asymptotic skewness $S$, kurtosis $K$, and ratio $R_2$ for nMBE models in $d=1+1$ (top) and $2+1$ (bottom). Data for $v_d=1/2$ in $2+1$ ES ICs.}
\begin{center}
\begin{tabular}{c c c c c c c c c c c}
\hline \hline
model   & & \multicolumn{3}{c}{flat}  & & \multicolumn{3}{c}{ES} & & $R_2$ \\
\hline
        & & $S$ & & $K$ & & $S$ & & $K$ & &  \\
\hline
CRSOS1  & &   0.137(8)   & &   -0.002(8)   & &  0.094(2)   & &   0.001(5)   & &   1.63(4)  \\
CRSOS4  & &   0.134(9)   & &   -0.001(1)   & &  0.090(2)   & &   0.000(1)   & &   1.62(5)  \\
DT      & &   0.136(8)   & &    0.001(1)   & &  0.093(4)   & &   0.001(2)   & &   1.69(6)  \\
\hline
CRSOS1  & &   0.13(2) & &   0.00(1)    & &  0.066(7)   & &   0.01(1)    & &   2.26(4)  \\
CRSOS2  & &   0.13(1)   & &   0.000(8)   & &  0.065(6)   & &   0.003(7)   & &   2.27(5)  \\
CRSOS4  & &   0.13(2)   & &   0.007(9)   & &  0.062(8)   & &   0.003(6)   & &   2.28(4)  \\
\hline\hline
\end{tabular}
\label{tab1}
\end{center}
\end{table}

Since the parameter $\Gamma$ is known only for the $1+1$ CRSOS1 model, to confirm the universality of the HDs, we investigate the (adimensional) cumulant ratios: skewness $S=\left\langle h^3 \right\rangle_c/\left\langle h^{2} \right\rangle_c^{3/2} \simeq \left\langle \chi^3 \right\rangle_c/\left\langle \chi^{2} \right\rangle_c^{3/2}$ and kurtosis $K=\left\langle h^4 \right\rangle_c/\left\langle h^{2} \right\rangle_c^2 \simeq \left\langle \chi^4 \right\rangle_c/\left\langle \chi^2 \right\rangle_c^2$. In the flat case, corrections $\mathcal{O}(t^{-\beta/2})$ and $\mathcal{O}(t^{-\beta})$ are found in $d=1+1$ and $2+1$, respectively (see Fig. \ref{fig1}c). For ES, the exponents seem consistent with twice the ones for flat IC, but the extrapolated values are almost the same if we assume identical corrections. The asymptotic values of $S$ and $K$ for all investigated models, in the same dimension and IC, agree quite well, as shown in Tab. \ref{tab1}, confirming the universality of the HDs, as well as their IC dependence. Interestingly, $K$ is always very close to zero. Moreover, for flat ICs, $S$ is almost the same for $1+1$ and $2+1$, so that these HDs have quite similar shapes, while in the ES case a decreasing $S$ is observed. This contrasts with the KPZ HDs, whose $S$ and $K$ are increasing functions of $d$ \cite{Alves14}, and it is possibly related to the fact that the nonlinearity in nMBE growth becomes irrelevant at its upper critical dimension $d_u=4$ \cite{Barabasi}, where $S$ and $K$ are expected to vanish. We recall that the corresponding values of $|S|$ and $|K|$ for KPZ HDs (with flat and ES ICs in $1+1$ and $2+1$) fall into the ranges $0.22 \lesssim |S| \lesssim 0.43$ and $0.09 \lesssim |K| \lesssim 0.35$ \cite{PraSpo1,healy12,Oliveira13}, being considerably larger than the ones in Tab. \ref{tab1}. Larger ratios ($|S| \approx 0.32$ and $|K| \approx 0.1$ in $d=1+1$ and $|S| \approx 0.20$ in $d=2+1$) have also been reported for the steady state HDs of the CRSOS model \cite{FabioVLDSold,*Tiagorug07}, while a much smaller skewness ($|S| \approx 0.0441$) was recently found in a (one-loop) renormalization analysis of the nMBE equation in this regime \cite{Tapas16}.

\begin{figure}[!t]
\includegraphics[height=3.52cm]{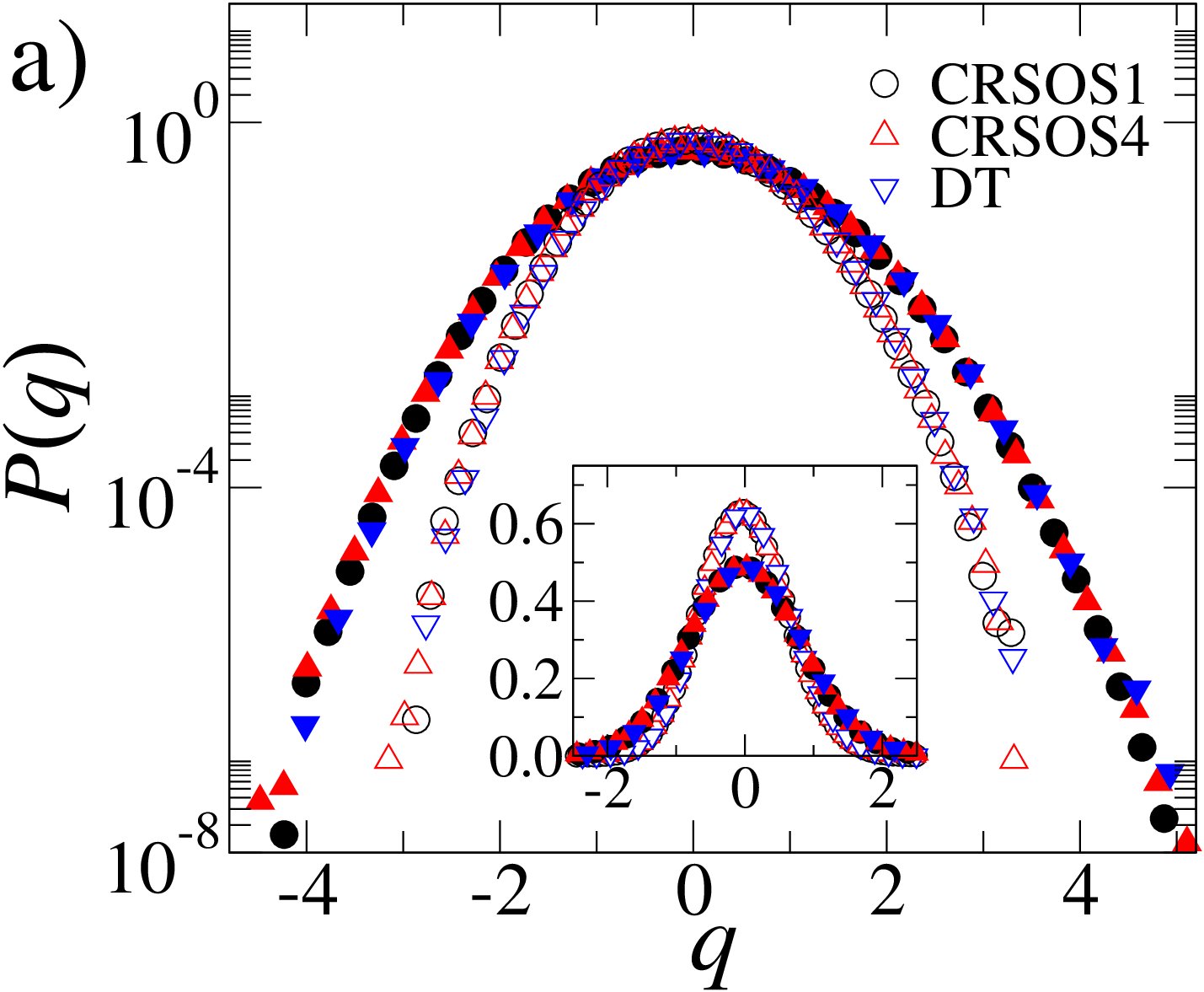}
\includegraphics[height=3.52cm]{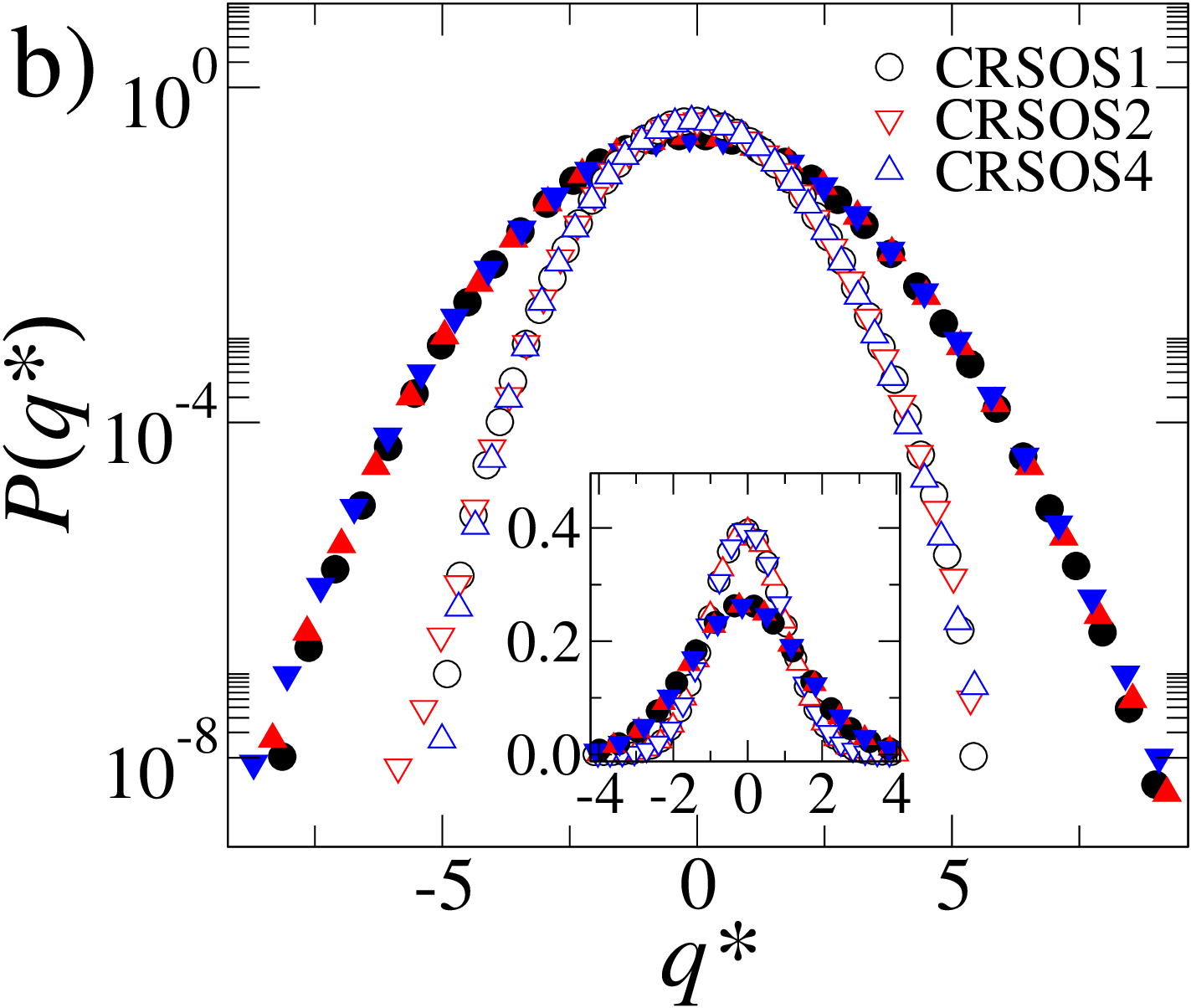}
\caption{(Color online) Rescaled HDs for models in a) $d=1+1$ and b) $2+1$. Distributions for flat (open - inner) and ES (solid symbols - outer curves, with $v_d=2$ in $2+1$) ICs are shown. Insets show the same data in linear scale.}
\label{fig2}
\end{figure}

Although, without knowing $\Gamma$, we cannot determine $\left\langle \chi^2 \right\rangle_c$ for all models, the product $g_2 \equiv \Gamma^{2\beta} \left\langle \chi^2 \right\rangle_c = \left\langle h^2 \right\rangle_c /t^{2\beta} + \ldots$ can be estimated, as done in Fig. \ref{fig1}d. Then, assuming the universality of the $\left\langle \chi^2 \right\rangle_c$'s in Tab. \ref{tab0}, one readily obtains $\Gamma= \left( g_2/\left\langle \chi^2 \right\rangle_c\right)^{1/2\beta} = 2.7(1)$ (CRSOS4) and $\Gamma= 0.035(2)$ (DT, with $N=20$) in one dimension. The reliability of such estimates is confirmed by the nice data collapse shown in Fig. \ref{fig2}a, where the HDs $P(q)$, with $q \equiv (h-t)/(\Gamma t)^{\beta}$, for different models are compared. We remark that these collapses confirm that $\Gamma$ is the same for fixed-size and enlarging substrates. Additional evidence of this is provided by the universality of the ``cross-subclass'' \cite{healy13} variance ratios $R_2 \equiv g_2^{ES}/g_2^{f} \simeq \left\langle \chi^2 \right\rangle_c^{ES}/\left\langle \chi^2 \right\rangle_c^{f}$, as shown in Tab. \ref{tab1}. To compare the $2+1$ HDs, we use the variable $q^* \equiv (h-t)/(\sqrt{g_2^{f}} t^{\beta})$, which turns out to be simply $q^*=q/\sqrt{\left\langle \chi^2 \right\rangle_c^{f}}$, so that flat and ES $P(q^*)$'s have variances 1 and $\left\langle \chi^2 \right\rangle_c^{ES}/\left\langle \chi^2 \right\rangle_c^{f}$, respectively. Again, a very good collapse is found (see Fig. \ref{fig2}b), which confirms that $2+1$ HDs are also universal and IC-dependent.

\begin{figure}[!b]
\includegraphics[width=4.25cm]{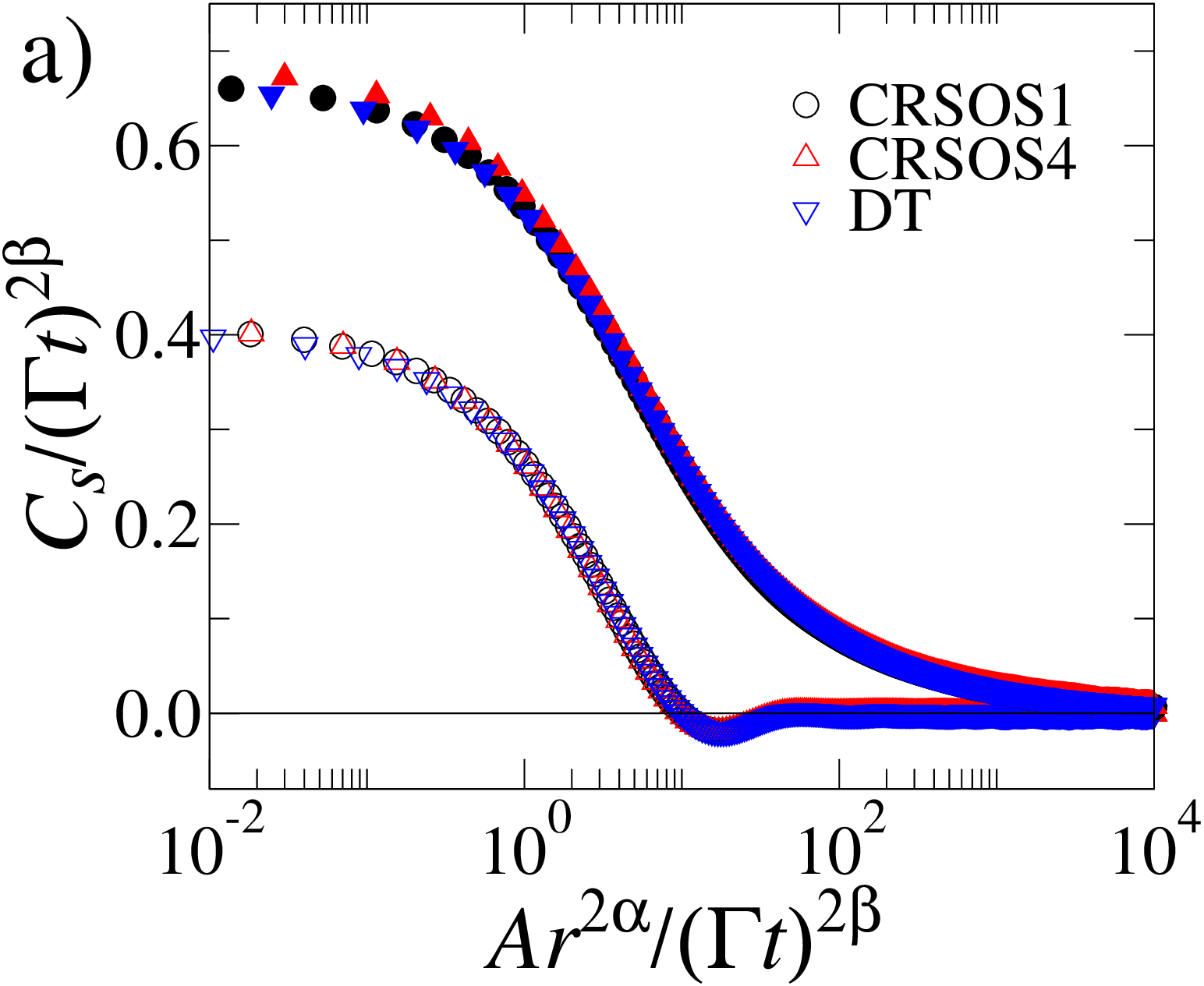}
\includegraphics[width=4.25cm]{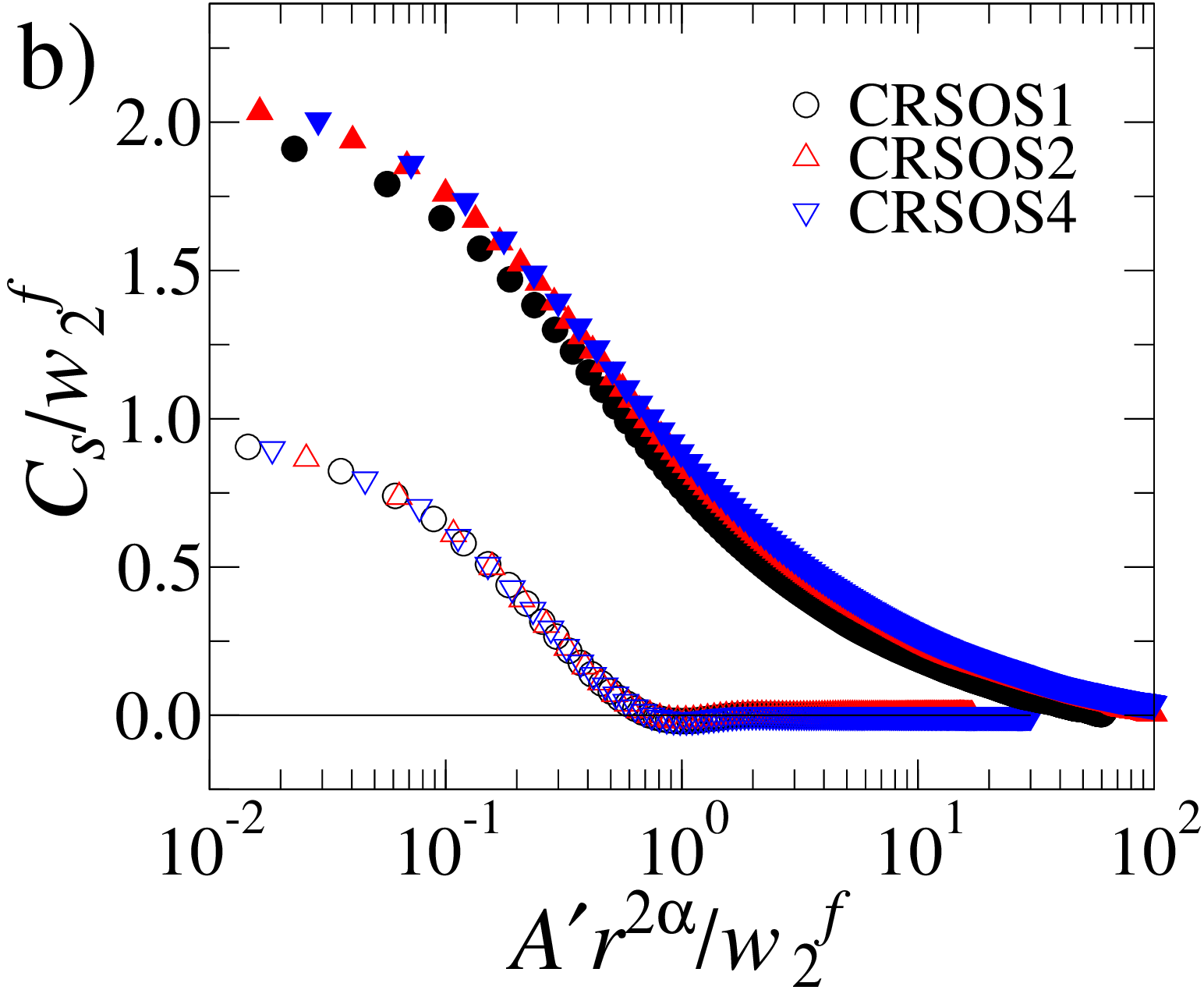}
\caption{(Color online) Rescaled spatial covariances for models with flat (open - bottom) and ES (solid symbols - top) ICs, in a) $d=1+1$ and b) $2+1$ (with $v_d=2$ in ES ICs).}
\label{fig3}
\end{figure}

Now, we turn to the analysis of the spatial covariance
\begin{equation}
 C_s(r,t) = \left\langle \tilde{h}(x,t) \tilde{h}(x+r,t) \right\rangle \simeq (\Gamma
t)^{2 \beta} \Psi[A r^{2\alpha}/(\Gamma t)^{2 \beta}],
\end{equation}
where $\tilde{h} \equiv h- \left\langle h \right\rangle$, $\Psi$ is a scaling function and $A$ is the same as defined above, in $d=1+1$. Figures~\ref{fig3}a and \ref{fig3}b show the rescaled $C_s$ for all investigated models in $d=1+1$ and $2+1$, respectively. Interestingly, the curves for flat ICs cross the zero and have a minimum in the negative region, indicating the existence of a characteristic length in the interfaces, which is not present when the substrate expands. Since the $A$'s are not known for the CRSOS4 and DT models (in $1+1$), we determine them by making the minima of their curves (in flat case) to coincide with the one for the CRSOS1 model. This yields $A = 8.67(8)$ (CRSOS4) and $A = 0.599(6)$ (DT). The constant $A'$ is obtained in the same way for $2+1$ curves, but shifting all minima to 1. Moreover, in this dimension one uses $w_2^{f}$ (obtained from simulations), instead of $(\Gamma t)^{2\beta}$ in the rescaling, so that $\Psi^{f}(0)=1$ and $\Psi^{ES}(0)=\left\langle \chi^2 \right\rangle_c^{ES}/\left\langle \chi^2 \right\rangle_c^{f}$. The good collapse of rescaled curves confirms that $\Psi^{f}(x)$ and $\Psi^{ES}(x)$ are universal, but $\Psi^{f}(x) \neq \Psi^{ES}(x)$. Hence, different processes exist for generating the flat and ES nMBE interfaces. It is worthy noting that the scaling functions [$\Psi(x)$'s] for $1+1$ and $2+1$ are very similar (for a given IC), when appropriately rescaled. For instance, in the ES subclass, one finds approximately $\Psi^{ES}(x) \sim x^{-1/2}$, for large $x$, in both dimensions.
 
As an aside, from estimates of $A$'s and $\Gamma$'s in $d=1+1$, one finds $D=\Gamma/A^{\frac{1}{2\beta}-1} \approx 0.89$ (CRSOS4) and $D \approx 0.046$ (DT, $N=20$). Moreover, disregarding the (small) two-loop correction in $A$, one obtains $\lambda_4 \approx \Gamma/A^{\frac{1}{2\beta}+\frac{1}{2}} \approx 0.03$ (CRSOS4) and $\lambda_4 \approx 0.098$ (DT, $N=20$).


\begin{figure}[!t]
\includegraphics[width=4.25cm]{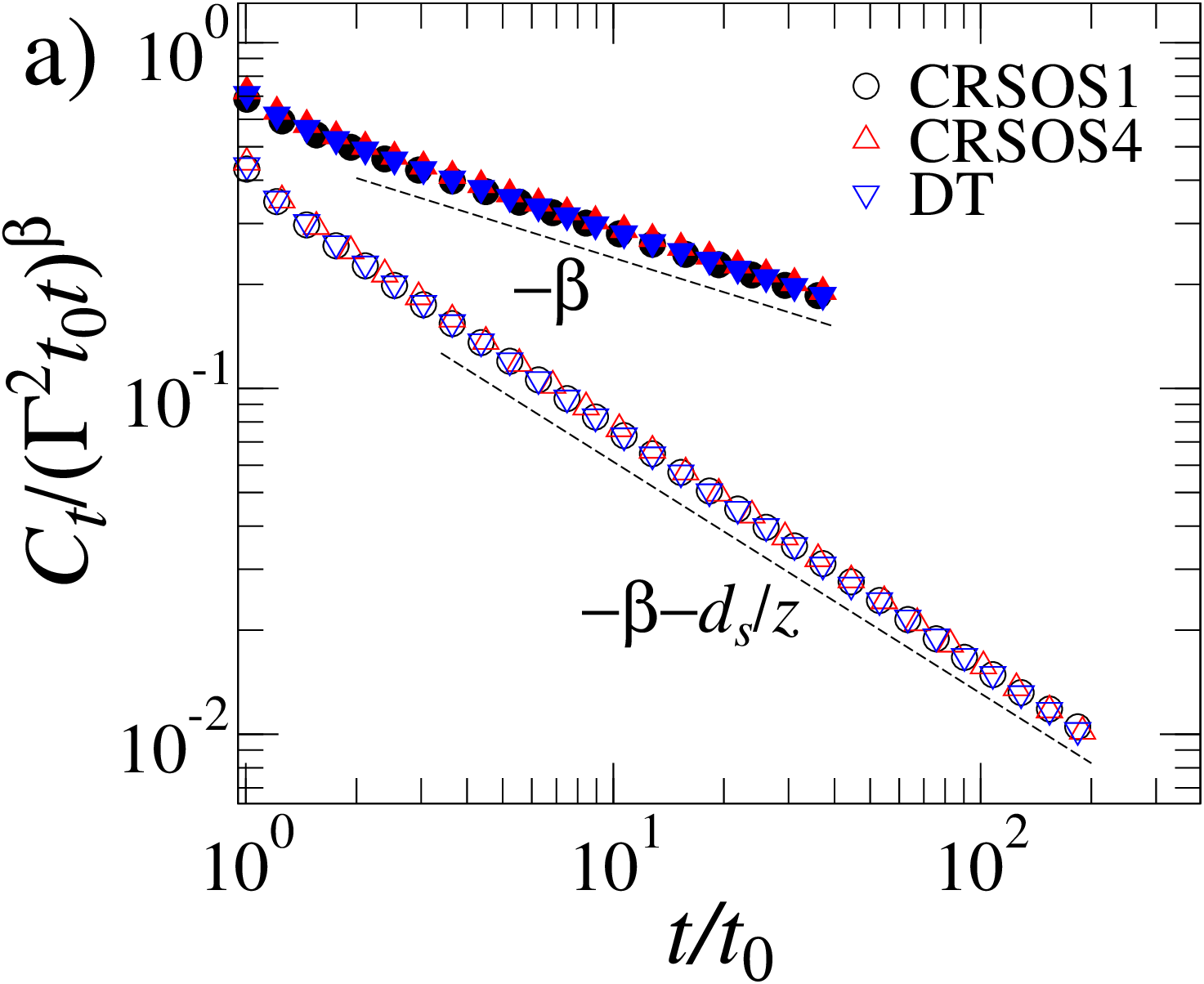}
\includegraphics[width=4.25cm]{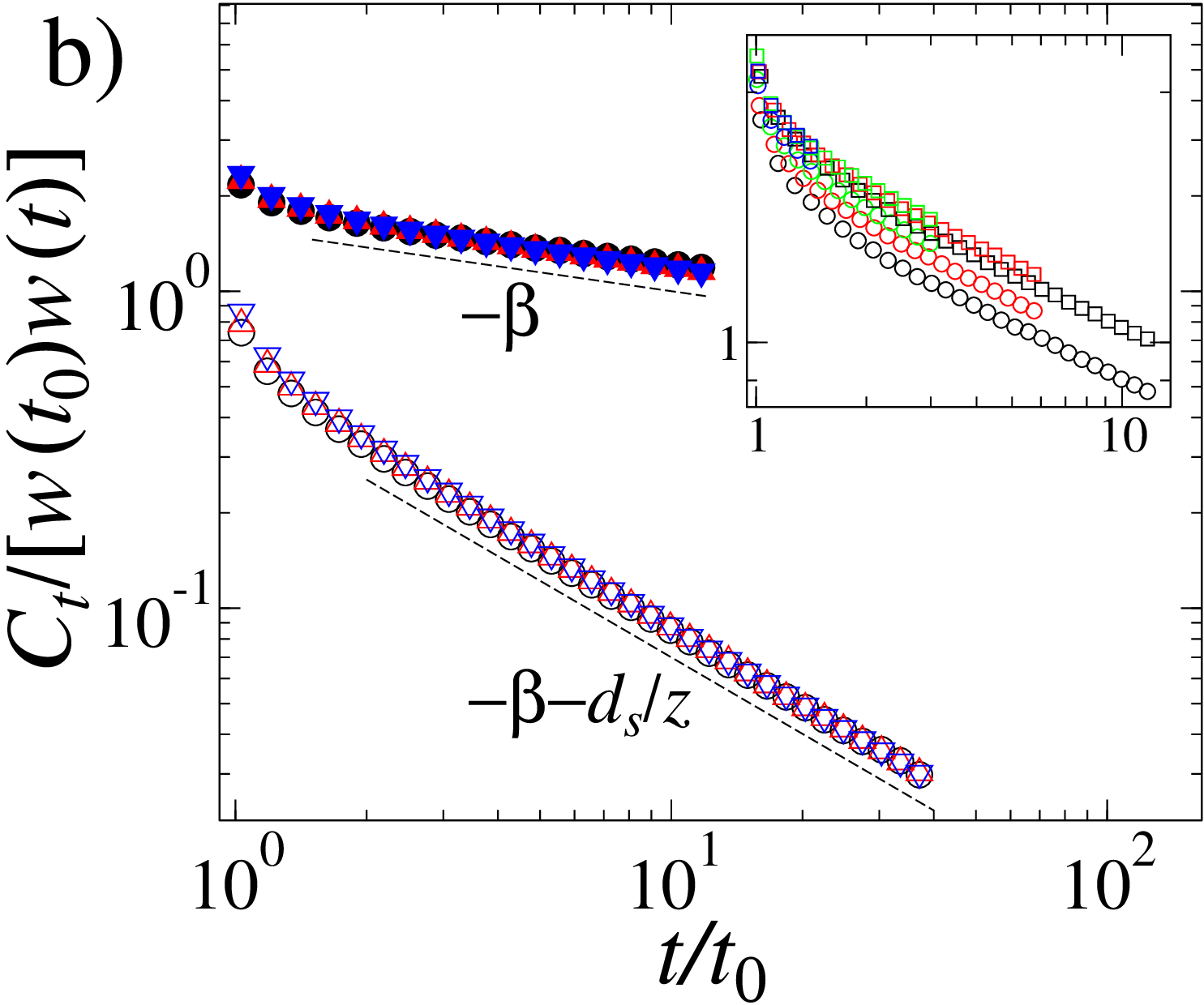}
\caption{(Color online) Rescaled temporal covariances for flat (open - bottom) and ES (solid symbols - top) ICs, in a) $1+1$ and b) $2+1$ (with $v_d=2$ for ES ICs). Dashed lines have the indicated slopes. The inset shows non-extrapolated curves for the ES case, in $d=2+1$, for different $t_0 \in [125,1000]$.}
\label{fig4}
\end{figure}

We also investigate the temporal covariance
\begin{equation}
 C_t(t,t_0) = \left\langle \tilde{h}(x,t_0) \tilde{h}(x,t) \right\rangle \simeq (\Gamma^2 t_0 t)^{\beta} \Phi(t/t_0).
\end{equation}
Once more, the nice data collapse displayed in Fig. \ref{fig4} demonstrates that universal IC-dependent scaling functions $\Phi(x)$ exist in the nMBE class. In $d=2+1$, we have used $w(t)$, rather than $(\Gamma t)^{\beta}$ in rescaling. Only data for $2+1$ ES ICs do not collapse well [see the inset of Fig. \ref{fig4}b], due to strong finite-time corrections $\mathcal{O}(t_0^{-2\beta})$, but when extrapolating the (rescaled) curves to $t_0 \rightarrow \infty$, a very good agreement is obtained, as the main plot of Fig. \ref{fig4}b shows. A similar procedure has been employed to analyze the universality of $\Phi(x)$ in the KPZ class \cite{Ismael14}. Substantially, in both dimensions, we find a power law decay $\Phi(x) \sim x^{-\bar{\lambda}}$, with exponents $\bar{\lambda}=\beta+d_s/z$ (flat) and $\bar{\lambda}=\beta$ (ES), in striking agreement with conjectures by Kallabis and Krug \cite{Kallabis99} and Singha \cite{Singha-JSM2005}, respectively.


In summary, we have demonstrated that 1-point height fluctuations in the nMBE class evolve, in the growth regime, according to the ``KPZ ansatz'' (Eq. \ref{eqansatz}) with universal and IC-dependent HDs. Moreover, 2-point spatial and temporal correlators are also IC-dependent. Therefore, the nMBE class splits into subclasses sharing the same critical exponents, similarly to KPZ systems. The absence of such splitting in HDs of linear classes, which are Gaussian for flat and ES ICs~\footnote {We have confirmed this by numerically integrating Eqs. \ref{eqKPZ} and \ref{eqVLDS} with $\lambda_2 = \lambda_4 = 0$ in $d=1+1$}, suggests that this is a feature of nonlinear interfaces, possibly due to the lack of an up-down reflection symmetry in them. We claim that our findings will be very useful to confirm the universality class of growing systems, along the same lines of Refs. \cite{TakeuchiPRL,*TakeuchiSP,Almeida14,*healy2014,*Almeida15,Iuri15}, especially because effective \textit{local} roughness exponents close to the nMBE value ($\alpha \approx 2/3$) have been found in grained/mounded films \cite{Barabasi,TiagoGraos1,*TiagoGraos2}, but they can be a simple consequence of a geometric effect \cite{TiagoGraos1,*TiagoGraos2}. From a theoretical side, our results will certainly motivate and guide analytical works toward exact solutions of the nMBE equation and related discrete models.

\acknowledgments

We acknowledge support from CNPq, CAPES and FAPEMIG (Brazilian agencies), and thank S. O. Ferreira for helpful discussions. T.J.O. appreciates the kind hospitality of the group of Prof. James Evans at Iowa State University, where part of this work was done.

\bibliography{HDsVLDS}

\end{document}